\documentclass[aps,prl,reprint,superscriptaddress]{revtex4-2}
\usepackage{amsfonts}
\usepackage{amssymb}
\usepackage{amsmath}
\usepackage{graphicx}
\usepackage{epsfig}
\usepackage{array}
\usepackage{braket}
\usepackage{multirow}
\usepackage[table,xcdraw]{xcolor}
\usepackage[colorlinks]{hyperref}

\begin{document}

\title{Chiral excitonic systems in twisted bilayers from F\"{o}rster
coupling and unconventional excitonic Hall effects}
\affiliation{School of Physics and Electronics, Hunan University, Changsha
410082, China} 
\affiliation{New Cornerstone Science Laboratory, Department
of Physics, University of Hong Kong, Hong Kong, China} 
\affiliation{HKU-UCAS
Joint Institute of Theoretical and Computational Physics at Hong Kong, China}
\author{Ci Li}
\email{lici@hnu.edu.cn}
\affiliation{School of Physics and Electronics, Hunan University, Changsha
410082, China}
\author{Wang Yao}
\email{wangyao@hku.hk}
\affiliation{New Cornerstone Science Laboratory, Department of Physics,
University of Hong Kong, Hong Kong, China} 
\affiliation{HKU-UCAS Joint
Institute of Theoretical and Computational Physics at Hong Kong, China}

\begin{abstract}
In twisted bilayer semiconductors with arbitrary twisting angles, a chiral
excitonic system can arise from the interlayer electron-hole Coulomb
exchange interaction (F\"{o}rster coupling) that hybridizes the anisotropic
intralayer excitons from individual layers. We present a general framework
for the effective exciton Hamiltonian taking into account the electron-hole
Coulomb exchange, using twisted homobilayer systems composed of transition
metal dichalcogenides or black phosphorus as examples. We demonstrate that such chiral excitonic systems can
feature unconventional Hall (Nernst) effects arising from quantum geometric
properties characteristic of the layer hybridized wavefunctions under the
chiral symmetry, for example, the time-reversal even layer Hall counter flow
and the crossed nonlinear dynamical Hall effect, when mechanical and
statistical force (temperature or density gradient) drives the exciton flow.
\end{abstract}

\maketitle





Layered two-dimensional (2D) semiconductors have provided a platform for the
various frontiers of condensed matter physics \cite%
{Nov,Wang1,Zhang,Gang,Kai,Mac1}. In particular, they offer exciting
opportunities to explore exciton physics and strong light-matter
interactions \cite{Wang1,Zhang,Gang}. In transition metal dichalcogenides
(TMDs), tightly bound Wannier excitons are formed around the degenerate $\pm
K$ valleys at the corners of the Brillouin zone (BZ), where optical
selection rules make possible manipulation of the valley pseudospin \cite%
{Wang2,Mak,Zen,Cao,Jon,Wan,Wang3,Urb,Hei,Hao}. 
The small Bohr radius $\sim O(1)$ nm~\cite{Lou1,Rei,Cro} underlies a
significant electron-hole (e-h) Coulomb exchange at finite center-of-mass
(COM) momentum \cite{Wang3,Wu,Yu,Song}. This effective coupling between the
exciton's valley pseudospin and COM degrees of freedom results in the
splitting of the exciton dispersion into two branches having linearly
polarized optical dipoles, longitudinal ($L$) and transverse ($T$) to
exciton momentum, respectively \cite{Wang3,Yu,Lou,Mac,Thy,Abr}. Recent
studies have found exotic properties of excitons in monolayer TMDs on
patterned substrates \cite{Yang1,Yang2} due to the massless $L$ branch's
sensitive dependence on the surrounding dielectric \cite{Mal,Shan}.
Additionally, applying strain to monolayer TMDs can break the rotational
symmetry. This symmetry breaking results in a non-zero contribution from
short-range e-h exchange interactions between excitons in different valleys,
leading to anisotropy and the formation of linearly dispersing Dirac saddle
points within the light cone \cite{Wang3,Yu}.

Black phosphorus (BP) is an example of highly anisotropic 2D semiconductors.
The monolayer has a direct bandgap of $1\sim 2\,\mathrm{eV}$ at the $\Gamma $
point in the BZ \cite{Neto,Gom,Wei}, suitable for hosting field-effect
transistors as well \cite{YbZ}. The highly anisotropic band structure of the
band edges leads to anisotropic intralayer excitons \cite{Gom,LYang}, which
exhibit linearly polarized optical dipoles, in contrast to the circularly
polarized one for isotropic valley excitons in monolayer TMDs \cite%
{Yu,Song,Lou}.

In layered structures, exciton's layer degree of freedom can also be
explored for novel functionalities, which is typically introduced by
interlayer carrier tunneling that leads to the hybridization of intra- and
inter-layer exciton species \cite{Zhang,Gang,Yu,Yu1}. On the other hand, F%
\"{o}rster coupling \cite{For}, i.e., the e-h Coulomb exchange, provides
another channel to introduce nontrivial layer structures for excitons. This
interaction can non-locally transfer an intralayer exciton between different
layers \cite{Seo,Kno}, enabling exciton propagation in the out-of-plane
direction. Its long-range nature allows it to function even when carrier
interlayer tunneling is quenched by spacer layers. Previous studies have
shown that cross-dimensional valley excitons can arise from this coupling in
arbitrarily twisted stacks of monolayer semiconductors \cite{Li,Li1}.

Like electrons where the band geometric quantities have been extensively
explored in transport phenomena \cite{Nag}, excitons can also have such
properties from the dependence of its internal degree of freedom on the COM
momentum, giving rise to novel exciton transport driven by
the external mechanical and statistical force (temperature or density
gradient) \cite{Wang,Ong}. In layered structures, this possibility can be
rooted in the layer degree of freedom in the exciton wavefunction which can
be engineered by twisting in the layered structure. Recently, two new types
of Hall effect have been discovered in chiral electronic systems in twisted
bilayers, i.e., the time-reversal even layer Hall counter flow (TREHCF) \cite%
{Zhai} and the crossed nonlinear dynamical Hall effect (CNDHE) \cite{Xiao,JC}%
. These findings uncover novel Hall physics in chiral electronic systems
that can be straightforwardly engineered by twisting. The question that
naturally arises is whether chiral excitonic systems can be similarly
engineered.

In this paper, we show that the structural chirality from twisting in a
layered structure can lead to a chiral exciton system when F\"{o}rster
coupling hybridizes intralayer excitons from the individual layers. In the
limit where carrier interlayer tunneling is quenched, we present a general
formulation of the effective exciton Hamiltonian taking into account the e-h
Coulomb exchange interactions, using twisted homobilayer TMDs and BP as
examples. The excitonic chirality requires the presence of
anisotropy that breaks the continuous in-plane rotational symmetry, which is
satisfied by the strong anisotropic exciton in BP. In the TMDs case, the
required anisotropy can be provided by either trigonal warping or strain,
and we considered the latter as an example. We demonstrate that such chiral
excitonic systems can also feature unconventional Hall (Nernst) effects
arising from quantum geometric properties characteristic of the layer
hybridized wavefunctions under the chiral symmetry, for example, the TREHCF
and the CNDHE, when mechanical and statistical force (temperature or density
gradient) drives the longitudinal exciton flow. These findings point to
exciton-based optoelectronic functionalities exploiting the layer quantum
degree of freedom in twisted structures.


\begin{figure*}[tbp]
\begin{center}
\includegraphics[width=0.75\textwidth]{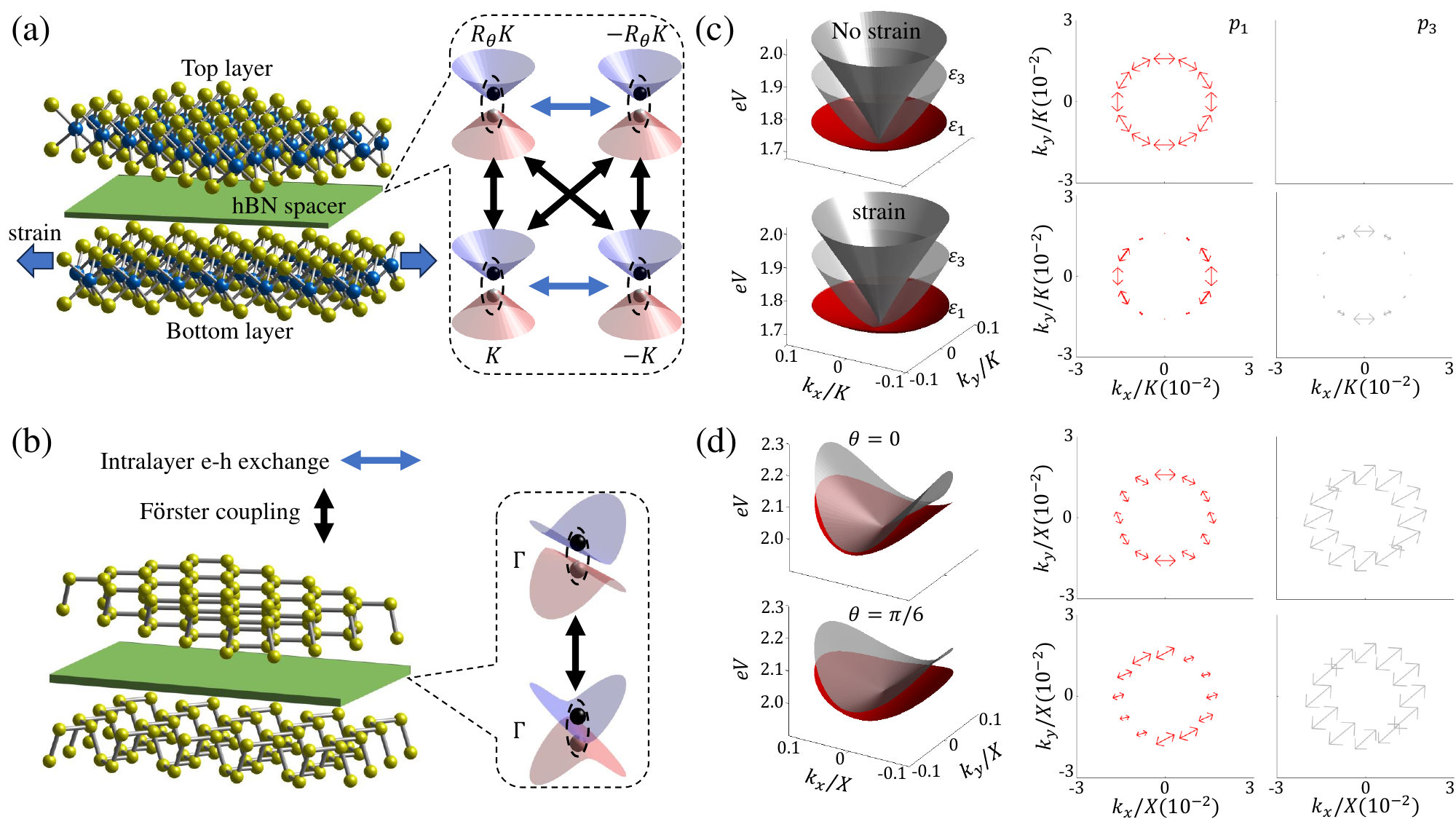}
\end{center}
\caption{(color online) (a) Left panel: Schematic of an arbitrarily twisted
homobilayer TMDs with strain (applied along $x$ direction with strain extent 
$\protect\epsilon _{st}\sim 1\%$ in the bottom layer), where hBN spacers
quench the charge hopping. Right panel: Illustration of electron-hole
Coulomb exchange between valley excitons in the same or different layers,
showing approximate band edges near high-symmetry points. (b) Similar plot
for twisted homobilayer BP. (c) Left panel: Excitonic dispersions for the
effective exciton Hamiltonian (\protect\ref{TMD}) of TMDs with and without
strain in untwisted situation. Right panel: Polarization distribution of the
optical dipole for two bands in momentum space. For clarity, only the
polarization distribution $p_{n}$ of the optical dipole for the marked band $%
\protect\varepsilon _{n}$ are shown. (d) Similar plots for the effective
exciton Hamiltonian (\protect\ref{BP}) of BP with different twisted angles $%
\protect\theta $.}
\label{fig1}
\end{figure*}

\section{Effective exciton Hamiltonians of twisted homobilayer TMDs and BP}

As previously stated, the effective exciton Hamiltonian in twisted bilayer
structures is primarily based on intralayer valley excitons in the two
layers. In the case of monolayer situations under the low-energy
approximation, there are typically at least two non-equivalent valleys
connected by rotational or time-reversal symmetry in the first BZ, such as
the $\pm K$ valley in monolayer TMDs \cite{Wang1,Wang2} (Fig. \ref{fig1}%
(a)), resulting in a $4\times 4$ effective Hamiltonian for intralayer
excitons, as discussed below. However, when a single high symmetry point is
present, such as the $\Gamma $ point in monolayer BP (Fig. \ref{fig1}(b)),
the Hamiltonian reduces to a simple $2\times 2$ matrix. In the following, we
present a method for obtaining the effective exciton Hamiltonian through a
general calculation of the e-h Coulomb exchange interaction, and use it to
acquire the exciton Hamiltonian for these two materials.

\subsection{General description the effective exciton Hamiltonian}

The construction of this Hamiltonian relies on two types of interaction: the
e-h exchange interaction between intralayer excitons of the same or
different valleys in the same layer and the F\"{o}rster coupling that
connects intralayer excitons in different layers, as illustrated in Fig. \ref%
{fig1}. For a specific layer $l$ ($l=t,b$ in this paper, $t/b$ means
top/bottom layer, as shown in Fig. \ref{fig1}(a)), the basis of valley
excitons can be expressed as $\left( \left\vert l,\boldsymbol{k}%
\right\rangle _{\alpha },\left\vert l,\boldsymbol{k}\right\rangle _{\beta
}\right) $, $\alpha ,\beta $ means two different valley indexes with the
in-plane COM momentum $\boldsymbol{k}=\left( k\cos \varphi ,k\sin \varphi
\right) $, and generally the effective exciton Hamiltonian in this situation
can be expressed as%
\begin{equation}
H_{\mathrm{ex}}=\left( 
\begin{array}{cc}
H_{k}^{b} & 0 \\ 
0 & H_{k}^{t}%
\end{array}%
\right) +\left( 
\begin{array}{cc}
H_{\mathrm{intra}}^{b} & H_{\mathrm{inter}}^{b,t} \\ 
H_{\mathrm{inter}}^{t,b} & H_{\mathrm{intra}}^{t}%
\end{array}%
\right) .  \label{Gen}
\end{equation}%
The first term is the kinetic energy of excitons in different layers, which
has the form%
\begin{equation}
H_{k}^{l}=\left( 
\begin{array}{cc}
\sum_{\gamma }\hbar ^{2}k_{\gamma }^{2}/\left( 2m_{ex,\alpha }^{\gamma
,l}\right)  & 0 \\ 
0 & \sum_{\gamma }\hbar ^{2}k_{\gamma }^{2}/\left( 2m_{ex,\beta }^{\gamma
,l}\right) 
\end{array}%
\right) ,
\end{equation}%
with $m_{ex,\alpha }^{\gamma ,l},\gamma =x,y$ is the exciton mass in $l$
layer, depending on the twisted angle $\theta _{l}$ of each layer (see
Supplementary \cite{Supp} for more details). 
\begin{eqnarray}
&&\left( 
\begin{array}{c}
\left\langle l,\boldsymbol{k}\right\vert _{\alpha } \\ 
\left\langle l,\boldsymbol{k}\right\vert _{\beta }%
\end{array}%
\right) H_{\mathrm{intra}}^{l}\left( \left\vert l,\boldsymbol{k}%
\right\rangle _{\alpha },\left\vert l,\boldsymbol{k}\right\rangle _{\beta
}\right)   \notag \\
&\equiv &\left( 
\begin{array}{cc}
J_{\alpha ,\alpha }^{l} & J_{\alpha ,\beta }^{l} \\ 
J_{\beta ,\alpha }^{l} & J_{\beta ,\beta }^{l}%
\end{array}%
\right) ,
\end{eqnarray}%
represents the e-h intra- or intervalley exchange between the excitons in
the same or different valleys in the layer $l$, where%
\begin{eqnarray}
J_{\lambda ,\lambda ^{\prime }}^{l} &\approx &\left[ \psi _{\lambda
}^{l}\left( 0\right) \right] ^{\ast }\psi _{\lambda ^{\prime }}^{l}\left(
0\right) \frac{V\left( \boldsymbol{k}\right) }{4}\times  \\
&&\left( k_{+}e^{-i\theta _{l}}d_{cv,\lambda }^{-}+k_{-}e^{i\theta
_{l}}d_{cv,\lambda }^{+}\right) \times   \notag \\
&&\left( k_{+}e^{-i\theta _{l}}d_{cv,\lambda ^{\prime }}^{-}+k_{-}e^{i\theta
_{l}}d_{cv,\lambda ^{\prime }}^{+}\right) ^{\ast },  \notag
\end{eqnarray}%
with $k_{\pm }=k_{x}\pm ik_{y}$, $d_{cv,\lambda }^{\pm }=d_{cv,\lambda
}^{x}\pm id_{cv,\lambda }^{y}$. $d_{cv,\lambda /\lambda ^{\prime }}^{\gamma
},\lambda ,\lambda ^{\prime }=\alpha ,\beta $ is the optical transition
dipole in $\gamma $ direction, between conduction ($c$) and valence ($v$)
band edges of $\lambda $ or $\lambda ^{\prime }$ valley. $\psi _{\lambda
}^{l}\left( 0\right) \sim 1/\sqrt{a_{B,\lambda }^{x,l}a_{B,\lambda }^{y,l}}$
with the exciton Bohr radius $a_{B,\lambda }^{\gamma ,l}$ for $\lambda $
valley can be seen as the square root of the probability for electron and
hole to overlap in an exciton. $V\left( \boldsymbol{k}\right) =2\pi
e^{2}/\left( \epsilon k\right) $ is the unscreened form for the Coulomb
potential with $\epsilon \equiv 4\pi \epsilon _{0}\epsilon _{r}$ \cite%
{Wang3,Yu,Lou}.%
\begin{eqnarray}
&&\left( 
\begin{array}{c}
\left\langle l,\boldsymbol{k}\right\vert _{\alpha } \\ 
\left\langle l,\boldsymbol{k}\right\vert _{\beta }%
\end{array}%
\right) H_{\mathrm{inter}}^{l,l^{\prime }}\left( \left\vert l^{\prime },%
\boldsymbol{k}\right\rangle _{\alpha },\left\vert l^{\prime },\boldsymbol{k}%
\right\rangle _{\beta }\right)  \\
&\equiv &\left( 
\begin{array}{cc}
J_{\alpha ,\alpha }^{l,l^{\prime }} & J_{\alpha ,\beta }^{l,l^{\prime }} \\ 
J_{\beta ,\alpha }^{l,l^{\prime }} & J_{\beta ,\beta }^{l,l^{\prime }}%
\end{array}%
\right) ,  \notag
\end{eqnarray}%
shows the F\"{o}rster coupling between layer $l$ and $l^{\prime }$, where%
\begin{eqnarray}
J_{\lambda ,\lambda ^{\prime }}^{l,l^{\prime }} &\approx &\psi _{\lambda
}^{l}\left( 0\right) \left[ \psi _{\lambda ^{\prime }}^{l^{\prime }}\left(
0\right) \right] ^{\ast }\frac{V\left( \boldsymbol{k},\Delta z\right) }{4}%
\times  \\
&&\left( k_{+}e^{-i\theta _{l}}d_{cv,\lambda }^{-}+k_{-}e^{i\theta
_{l}}d_{cv,\lambda }^{+}\right) \times   \notag \\
&&\left( k_{+}e^{-i\theta _{l^{\prime }}}d_{cv,\lambda ^{\prime
}}^{-}+k_{-}e^{i\theta _{l^{\prime }}}d_{cv,\lambda ^{\prime }}^{+}\right)
^{\ast }.  \notag
\end{eqnarray}%
Here the Coulomb potential $V\left( \boldsymbol{k},\Delta z\right) =2\pi
e^{2}/\left( \epsilon k\right) \times \exp \left( -k\Delta z\right) $ is
related with the interlayer distance $\Delta z$ between two layers \cite{Li}%
. In the rest of this paper, we take $\Delta z=1\,\mathrm{nm}$ as a typical
value. Since the optical transition dipole $d_{cv,\lambda }^{\gamma }$ can
be got from the effective single-particle two-band $\boldsymbol{k}\cdot 
\boldsymbol{p}$ model near $\lambda $ valley, also the exciton Bohr radius $%
a_{B,\lambda }^{\gamma ,l}$, the construction has been finished.

\subsection{The exciton Hamiltonian of TMDs}

For the TMD cases, there are two inequivalent valleys as $\alpha \equiv 
\boldsymbol{K}$ and $\beta \equiv -\boldsymbol{K}$, connected by the
time-reversal symmetry. It has been shown that the exciton dispersion is
isotropic for the first order of $\boldsymbol{k}$ in monolayer TMDs \cite%
{Wang3,Yu,Lou}, these two properties lead to $H_{k}^{l}=\frac{\hbar ^{2}k^{2}%
}{2m_{ex}}=H_{k}$ and $a_{B,\lambda }^{x,l}=a_{B,\lambda
}^{y,l}=a_{B,\lambda ^{\prime }}^{x,l^{\prime }}=a_{B,\lambda ^{\prime
}}^{y,l^{\prime }}=a_{B}$. Without loss of generality, we let $\theta _{b}=0$
and $\theta _{t}=\theta $ and replace the twist angle $\theta $ in momentum
space into the real space one, i.e., $\theta \rightarrow -\theta $ in the
rest of the paper. The effective Hamiltonian $H_{\mathrm{TMD}}$ of valley
excitons in the twisted homobilayer TMD can be written as%
\begin{equation}
H_{\mathrm{TMD}}=\frac{\hbar ^{2}k^{2}}{2m_{ex}}+\sum_{l=t,b}H_{\mathrm{intra%
}}^{l}+\sum_{l,l^{\prime }=t,b}H_{\mathrm{inter}}^{l,l^{\prime }},
\label{TMD}
\end{equation}%
in the basis $\left\{ \left\vert l,\boldsymbol{k}\right\rangle
_{K},\left\vert l,\boldsymbol{k}\right\rangle _{-K}\right\} $, with%
\begin{eqnarray*}
H_{\mathrm{intra}}^{b} &=&J\frac{k}{K}\left( 
\begin{array}{cc}
1 & -e^{-2i\varphi } \\ 
-e^{2i\varphi } & 1%
\end{array}%
\right) , \\
H_{\mathrm{intra}}^{t} &=&J\frac{k}{K}\left( 
\begin{array}{cc}
1 & -e^{-2i\left( \theta +\varphi \right) } \\ 
-e^{2i\left( \theta +\varphi \right) } & 1%
\end{array}%
\right) ,
\end{eqnarray*}%
and%
\begin{equation*}
H_{\mathrm{inter}}^{b,t}=\left( H_{\mathrm{inter}}^{t,b}\right) ^{\dagger }=J%
\frac{k}{K}e^{-k\Delta z}\left( 
\begin{array}{cc}
e^{i\theta } & -e^{-i\left( \theta +2\varphi \right) } \\ 
-e^{i\left( \theta +2\varphi \right) } & e^{-i\theta }%
\end{array}%
\right) ,
\end{equation*}%
giving the dispersion as \cite{Li}%
\begin{eqnarray}
\varepsilon _{1} &=&\varepsilon _{2}=\frac{\hbar ^{2}k^{2}}{2m_{ex}}, \\
\varepsilon _{3} &=&\frac{\hbar ^{2}k^{2}}{2m_{ex}}+2J\frac{k}{K}-2J\frac{k}{%
K}e^{-k\Delta z},  \notag \\
\varepsilon _{4} &=&\frac{\hbar ^{2}k^{2}}{2m_{ex}}+2J\frac{k}{K}+2J\frac{k}{%
K}e^{-k\Delta z},  \notag
\end{eqnarray}%
which is independent of the twisted angle $\theta $. $m_{ex}\approx m_{e}$
is the effective exciton mass in monolayer TMDs. Here $K=4\pi /3a$, $a$
being TMD's lattice constant, and $J\sim 1\,\mathrm{eV}$ can be extracted
from first principle wavefunctions and exciton spectrum \cite{Wang2,Yu,Lou}.
The isotropic properties of the exciton bands in momentum space are evident
from Fig. \ref{fig1}(c).

If there is a strain in the bottom layer of twisted bilayer TMDs, it can be
transferred into the top layer with some loss \cite{Amo,Kum,Hei2}. This
mechanism leads to chiral symmetric excitons in twisted bilayer TMDs, where
the short-range e-h exchange contributes to the coupling between excitons in
different valleys of the same layer. This results in the non-diagonal term
of $H_{\mathrm{intra}}^{l}$ becoming $J_{0}^{l}-J\frac{k}{K}e^{\pm i\left(
\theta _{l}+2\varphi \right) }$ \cite{Supp}. In our consideration, $%
J_{0}^{b}\approx -6\,\mathrm{meV}$ arises from the strain along $x$
direction with an extent of $\epsilon _{st}=1\%$ \cite{Yu}, and $%
J_{0}^{t}\approx J_{0}^{b}\frac{\cos 2\theta +\sin 2\theta }{2}$ is
twisted-angle dependent \cite{Supp}. The related exciton dispersions, with
and without strain in untwisted situation, are displayed in Fig. \ref{fig1}%
(c), which also shows the polarization of the optical dipole for the band $%
\varepsilon _{n=1,3}$. The anisotropic polarization distribution of optical
dipole in momentum space indicates that the excitonic chirality appears in
twisted bilayer TMDs when strain is applied.

\subsection{The exciton Hamiltonian of BP}

The exciton Hamiltonian of BP can be obtained in the basis of $\{\left\vert
b,\boldsymbol{k}\right\rangle _{\Gamma },\left\vert t,\boldsymbol{k}%
\right\rangle _{\Gamma }\}$ by using the same approach as for TMDs. Since
the band edge only appear in the $\Gamma $ point per layer under the
low-energy approximation, i.e., $\alpha =\beta \equiv \Gamma $, this
Hamiltonian takes a simple $2\times 2$ form as \cite{Supp} 
\begin{eqnarray}
H_{\mathrm{BP}} &=&\left( 
\begin{array}{cc}
\sum_{\gamma }\hbar ^{2}k_{\gamma }^{2}/\left( 2m_{ex}^{\gamma ,b}\right)  & 
0 \\ 
0 & \sum_{\gamma }\hbar ^{2}k_{\gamma }^{2}/\left( 2m_{ex}^{\gamma
,t}\right) 
\end{array}%
\right)   \notag \\
&&+\left( 
\begin{array}{cc}
J_{\Gamma }^{b} & J_{\Gamma }^{b,t} \\ 
J_{\Gamma }^{t,b} & J_{\Gamma }^{t}%
\end{array}%
\right) ,  \label{BP}
\end{eqnarray}%
The exciton dispersions of $H_{\mathrm{BP}}$ has the from%
\begin{eqnarray}
\varepsilon _{1,2} &=&\hbar ^{2}\left(
M_{x}^{+}k_{x}^{2}+M_{y}^{+}k_{y}^{2}\right) + \\
&&\frac{k}{X}\frac{\mathcal{J}^{b}\cos ^{2}\varphi +\mathcal{J}^{t}\cos
^{2}\left( \varphi +\theta \right) }{2}  \notag \\
&&\pm \sqrt{A^{2}+B^{2}},  \notag
\end{eqnarray}%
where $A=\hbar ^{2}\left( M_{x}^{-}k_{x}^{2}+M_{y}^{-}k_{y}^{2}\right) +%
\frac{k}{X}\frac{\mathcal{J}^{b}\cos ^{2}\varphi -\mathcal{J}^{t}\cos
^{2}\left( \varphi +\theta \right) }{2}$ and $B=\sqrt{\mathcal{J}^{b}%
\mathcal{J}^{t}}\frac{k}{X}e^{-kz}\cos \varphi \cos \left( \varphi +\theta
\right) $ are functions of the twist angle $\theta $, with%
\begin{equation*}
M_{x}^{\pm }=\frac{m_{ex}^{x,b}\pm m_{ex}^{x,t}}{4m_{ex}^{x,b}m_{ex}^{x,t}}%
,M_{y}^{\pm }=\frac{m_{ex}^{y,b}\pm m_{ex}^{x,t}}{4m_{ex}^{y,b}m_{ex}^{y,t}}.
\end{equation*}%
Unlike in TMDs, $\theta $ affects the excitonic properties of $H_{\mathrm{BP}%
}$, which can be seen from the highly anisotropic dispersion at $\theta =0$
and $\pi /6$, also the highly anisotropic polarization distribution of
optical dipole, as shown in Fig. \ref{fig1}(d).

\begin{figure*}[tbp]
\begin{center}
\includegraphics[width=0.75\textwidth]{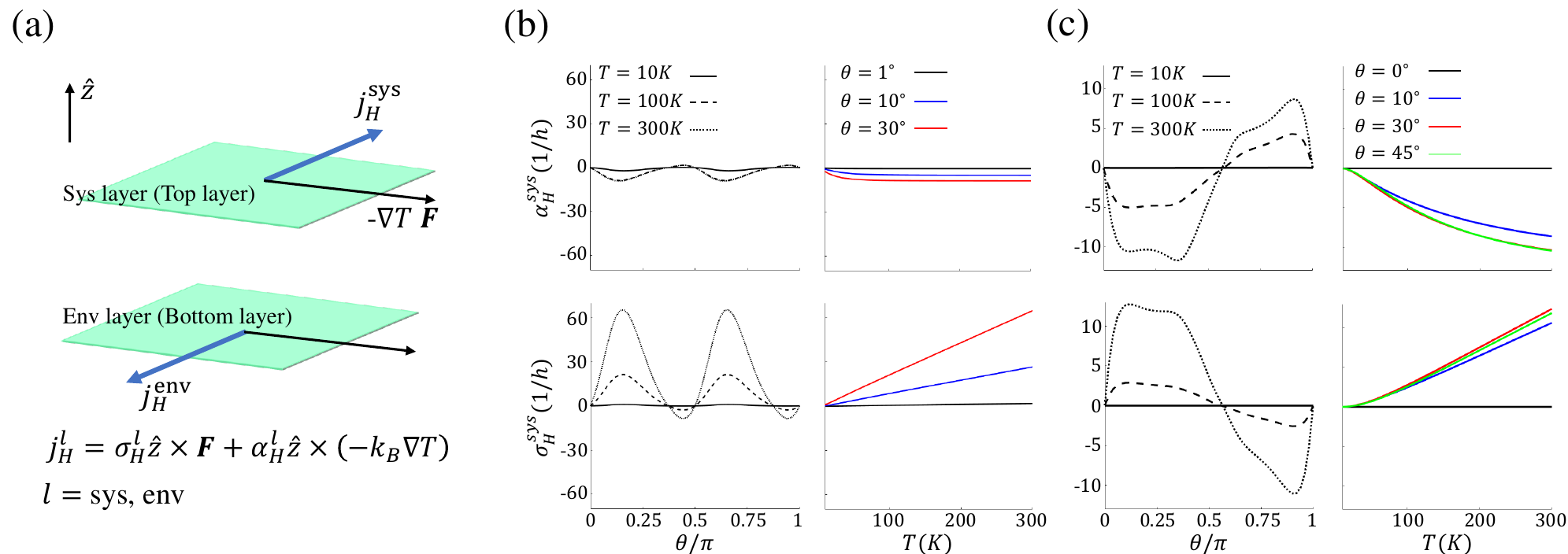}
\end{center}
\caption{(color online) (a) Schematic illustration of the time reversal even
layer Hall (Nernst) counter flow for excitons in twisted bilayer systems.
(b) Numerical calculation of $\protect\sigma _{H}^{\mathrm{sys}}$ and $%
\protect\alpha _{H}^{\mathrm{sys}}$ (Eq. \protect\ref{TRE}) for $H_{\mathrm{%
TMD}}$ (Eq. \protect\ref{TMD}) when strain is applied, with respect to the
twisted angle $\protect\theta $ and the temperature $T$, respectively. $%
\protect\mu \approx 1.69\,\mathrm{eV}$. (c) Similar plot for $H_{\mathrm{BP}}
$ (Eq. \protect\ref{BP}), with $\protect\mu \approx 2\,\mathrm{eV}$. $%
\protect\tau =1\,\mathrm{ps}$ for all plots \protect\cite{Zhai}.}
\label{fig2}
\end{figure*}

\section{The time-reversal even layer Hall (Nernst) counter flow and crossed
nonlinear dynamical Hall (Nernst) effect}

\subsection{The time-reversal even layer Hall (Nernst) counter flow (TREHCF)}

The non-zero TREHCF requires chiral symmetry in real space \cite{Zhai},
which is naturally satisfied by the twisted bilayer structure, as shown in
Fig. \ref{fig2}(a). In twisted bilayer chiral excitonic systems, this effect
can be defined as similar as the one for twisted bilayer electric systems 
\cite{Zhai}, which comes from the semi-classical calculation for the current
density of each layer%
\begin{equation*}
\boldsymbol{j}^{\mathrm{sys/env}}=\sum_{n}\int \frac{d^{2}\boldsymbol{k}}{%
\left( 2\pi \right) ^{2}}f_{n}\left( \boldsymbol{k}\right) \boldsymbol{v}%
_{n}^{\mathrm{sys/env}}\left( \boldsymbol{k}\right) ,
\end{equation*}%
where two layers are divided into the system layer (top layer) and
environment layer (bottom layer). $\boldsymbol{j}^{\mathrm{sys}}=-%
\boldsymbol{j}^{\mathrm{env}}$ preserves by the Onsager relation \cite{Zhai}%
. $f_{n}\left( \boldsymbol{k}\right) \approx f_{n}^{0}\left( \boldsymbol{k}%
\right) -\tau \frac{df_{n}^{0}\left( \boldsymbol{k}\right) }{dt}$ is the
off-equilibrium distribution function expanded in the first order of
relaxation time $\tau $, with $f_{n}^{0}\left( \boldsymbol{k}\right) =1/%
\left[ \exp \left( \frac{\varepsilon _{n}\left( \boldsymbol{k}\right) -\mu }{%
k_{B}T}\right) -1\right] $ is the equilibrium Bose-Einstein distribution
function. $\boldsymbol{v}_{n}^{\mathrm{sys/env}}\left( \boldsymbol{k}\right)
=\left\langle u_{n}\left( \boldsymbol{k}\right) \right\vert \frac{1}{2}%
\left\{ \boldsymbol{\hat{v}},\hat{P}^{\mathrm{sys/env}}\right\} \left\vert
u_{n}\left( \boldsymbol{k}\right) \right\rangle $ is the velocity projection
on system/environment layer of band $n$, where $\hat{P}^{\mathrm{sys/env}%
}=\left( 1\pm \hat{\sigma}_{z}\right) /2$ is the projection operator onto
the system/environment layer, with the Pauli matrix $\hat{\sigma}_{z}$ being
the out-of-plane layer-pseudospin operator operating in the layer index
subspace \cite{Hei1}. $\boldsymbol{\hat{v}}=\frac{\partial H_{\mathrm{ex}%
}\left( \boldsymbol{k}\right) }{\hbar \partial \boldsymbol{k}}$ is the total
velocity operator.

If we consider that there is an inhomogeneous distribution of the
temperature $T$ in real space and a general mechanical in-plane force $%
\boldsymbol{F}$ as $\boldsymbol{F}=\hbar \boldsymbol{\dot{k}}$. After some
tedious derivations \cite{Supp}, we can get the time-reversal even Hall
conductivity for the layer counter flow of the system as $\sigma _{H}^{%
\mathrm{sys}}=\tau \mathcal{V}/\hbar $, while $\alpha _{H}^{\mathrm{sys}%
}=\tau \mathcal{V}_{\mathrm{Ner}}/\hbar $ represents the time-reversal even
Nernst conductivity for the layer counter flow, where%
\begin{eqnarray}
\mathcal{V} &=&\sum_{n}\int \frac{d^{2}\boldsymbol{k}}{\left( 2\pi \right)
^{2}}f_{n}^{0}\omega _{n}\left( \boldsymbol{k}\right) ,  \label{TRE} \\
\mathcal{V}_{\mathrm{Ner}} &=&\sum_{n}\int \frac{d^{2}\boldsymbol{k}}{\left(
2\pi \right) ^{2}}\frac{\varepsilon _{n}-\mu }{k_{B}T}f_{n}^{0}\omega
_{n}\left( \boldsymbol{k}\right) ,  \notag
\end{eqnarray}%
with $\omega _{n}\left( \boldsymbol{k}\right) =\frac{1}{2}\left[ \nabla _{%
\boldsymbol{k}}\times \boldsymbol{v}_{n}^{\mathrm{sys}}\left( \boldsymbol{k}%
\right) \right] $ means the $\boldsymbol{k}$-space vorticity of the layer
current \cite{Zhai}, exhibiting the nontrivial quantum geometric properties
characteristic of the layer hybridized wavefunctions under the chiral
symmetry.

\begin{figure}[tbp]
\begin{center}
\includegraphics[width=0.45\textwidth]{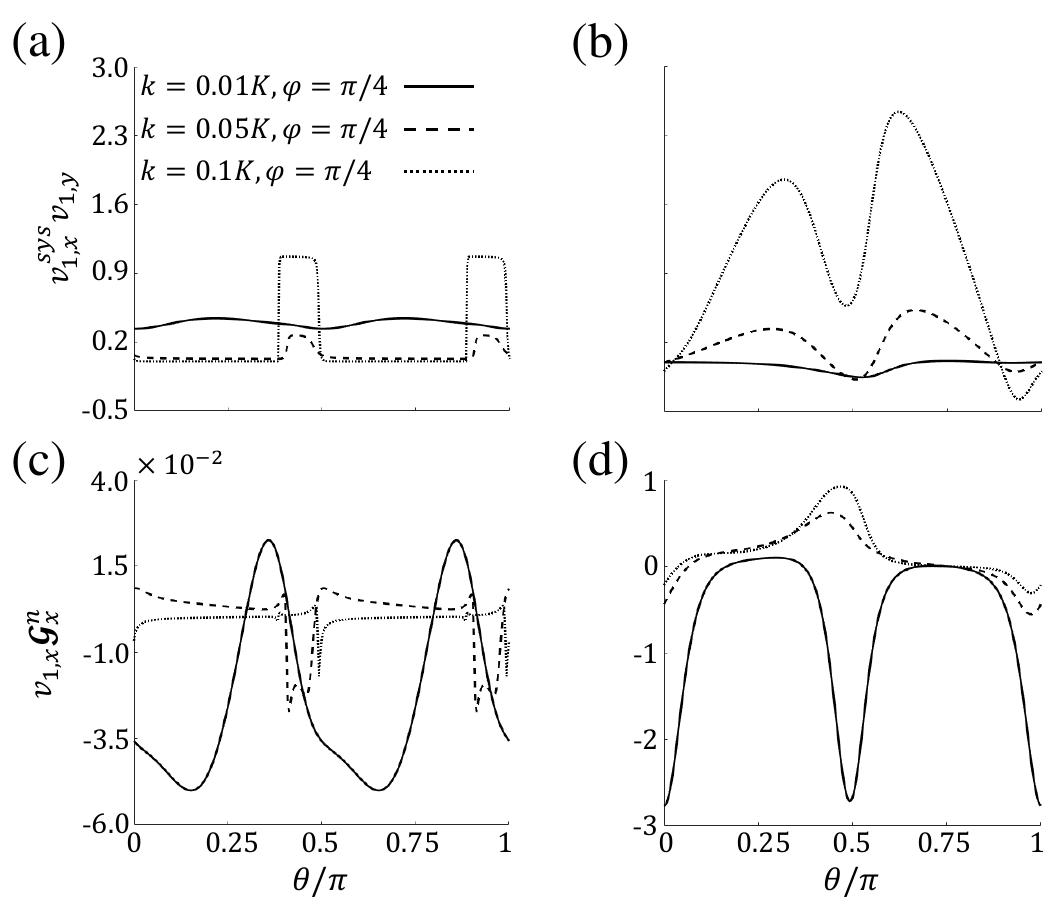}
\end{center}
\caption{(color online) (a) Numerical calculation of the product of the
velocity projection on system layer, $v_{n,x}^{sys}$, and the total band
velocity, $v_{n,y}$, in twisted bilayer TMDs with strain. $n=1$ represents
the ground states. All parameters used are the same as Fig. \protect\ref%
{fig2}(b). (b) Similar plot as (a) for BP. All parameters used are the same
as Fig. \protect\ref{fig2}(c). The unit on $y$-axis is ($\mathrm{\mathring{A}%
}/\mathrm{fs)}^{2}$ for both (a) and (b). (c) Numerical calculation of the
product of the total band velocity, $v_{1,x}$, and the interlayer Berry
connection polarizability, $\mathcal{G}_{y}^{1}\left( \boldsymbol{k}\right) $%
, in twisted bilayer TMDs with strain. All parameters used are the same as
Fig. \protect\ref{fig4}(b). (d) Similar plot as (c) for BP. All parameters
used are the same as Fig. \protect\ref{fig4}(c). The unit on $y$-axis is $%
\mathrm{\mathring{A}}^{2}\cdot \mathrm{fs}/\mathrm{eV}$ for both (c) and
(d). }
\label{fig3}
\end{figure}

In TMDs without strain, although the chiral symmetry in real space is
preserved, the rotational symmetry of the monolayer TMDs ensures that the
effective exciton Hamiltonian is independent of the twisted angle $\theta $ 
\cite{Li}. This reinstates the mirror symmetry and leads to isotropic
dispersions, as shown in Fig. \ref{fig1}(c). Consequently, the TREHCF is
forbidden. However, if strain is applied to the bottom layer of twisted
bilayer TMDs, strain transfer between the two layers \cite{Kum,Hei2} will
result in a twisted-angle-dependent strain in the top layer \cite%
{Kum,Hei2,Supp}. These two inequivalent strains in the different layers can
break the rotational symmetry in the exciton Hamiltonian, leading to the
excitonic chirality and a obvious TREHCF, as shown in Fig. \ref{fig2}(b).
Both $\sigma _{H}^{\mathrm{sys}}$ and $\alpha _{H}^{\mathrm{sys}}$ are
periodic functions of $\theta $, with a period of $\pi /2$ dominated by the
quantum geometric properties of $\omega _{n}\left( \boldsymbol{k}\right) $,
as shown from the product of two velocities in Fig. \ref{fig3}(a) \cite%
{Note1}.

In BP, due to the high anisotropy, the excitonic chirality is naturally
satisfied, leading to anisotropic exciton dispersions and a non-zero TREHCF,
as shown in Figs. \ref{fig1}(d) and \ref{fig2}(c). Here, $\sigma _{H}^{%
\mathrm{sys}}$ and $\alpha _{H}^{\mathrm{sys}}$ are also periodic functions
of $\theta $, but with a period of $\pi $ due to the $C_{2}$ symmetry of the
product of two velocities within the integral, as illustrated in Fig. \ref%
{fig3}(b). The chemical potential $\mu $ in Fig. \ref{fig2} is chosen such
that $\min \varepsilon _{1}-\mu \approx 0.005\,\mathrm{eV}$, where $%
\varepsilon _{1}$ is the ground energy of the exciton Hamiltonian, and $\min
\varepsilon _{1}$ in both materials is independent of $\theta $. Based on
this consideration, the changes of $\sigma _{H}^{\mathrm{sys}}$ and $\alpha
_{H}^{\mathrm{sys}}$ with respect to $T$ can be easily understand from the $T
$ dependence of $f_{n=1}^{0}$ since it obviously increases as $T$ rises.
This implies that the TREHCF for excitons in chiral bilayer structures could
be measurable at room temperature.

\begin{figure*}[tbp]
\begin{center}
\includegraphics[width=0.75\textwidth]{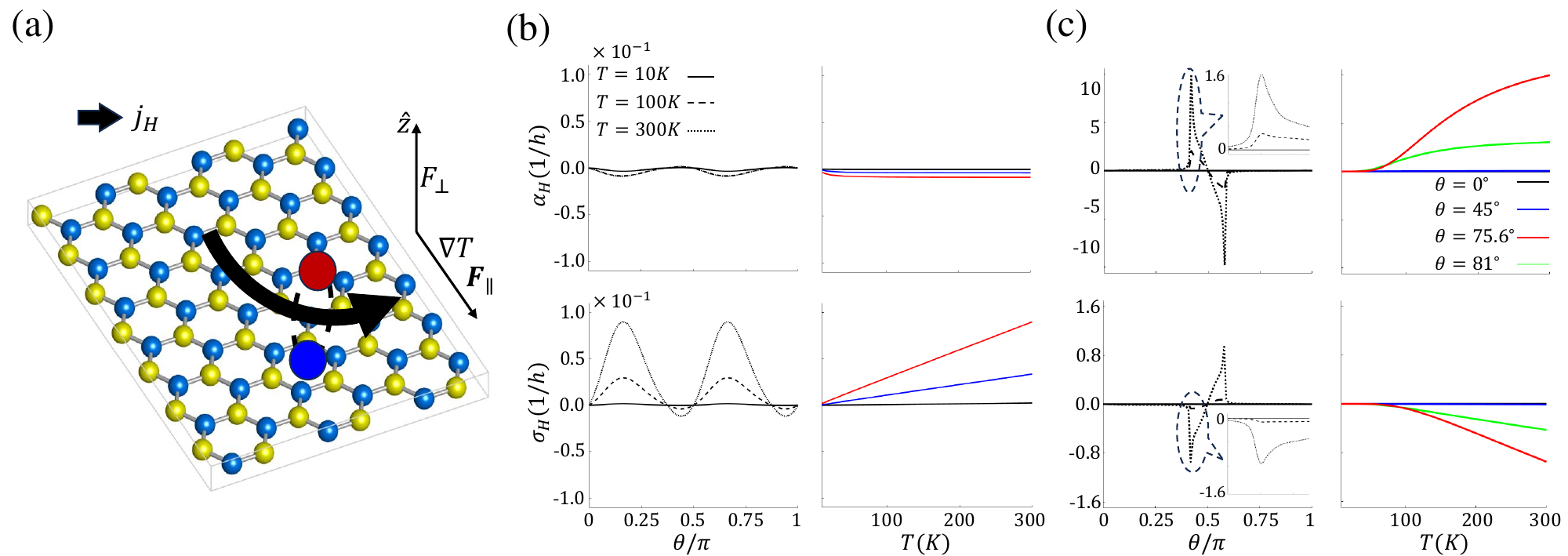}
\end{center}
\caption{(color online) (a) Schematic illustration of the crossed nonlinear
dynamical Hall (Nernst) effect for excitons in twisted bilayer systems. (b)
Numerical calculation of $\protect\sigma _{H}$ and $\protect\alpha _{H}$
(Eq. \protect\ref{CND}) for effective exciton Hamiltonian $H_{\mathrm{TMD}}$
(Eq. \protect\ref{TMD}) when strain is applied, with respect to the twisted
angle $\protect\theta $ and the temperature $T$, respectively. $\protect\mu %
\approx 1.69\,\mathrm{eV}$. (c) The similar plot for effective exciton
Hamiltonian $H_{\mathrm{BP}}$ (Eq. \protect\ref{BP}), with $\protect\mu %
\approx 1.98\,\mathrm{eV}$ with a gap $\protect\varepsilon _{d}=0.02\,%
\mathrm{eV}$ induced by the physical control of layers such as strain. $%
F_{\perp }^{0}=0.1\,\mathrm{eV/nm}$, $\protect\omega /2\protect\pi =0.1\,%
\mathrm{THz}$ for all plots \protect\cite{Xiao}.}
\label{fig4}
\end{figure*}

\subsection{The crossed nonlinear dynamical Hall (Nernst) effect (CNDHE)}

According to previous theoretical works \cite{Xiao,JC}, this effect can be
described from the similar starting point by calculating the current density
of excitons in the intrinsic response%
\begin{equation}
\boldsymbol{j}=\sum_{n}\int \left[ \frac{d^{2}\boldsymbol{k}}{\left( 2\pi
\right) ^{2}}f_{n}^{0}\left( \boldsymbol{k}\right) \boldsymbol{v}_{n}\left( 
\boldsymbol{k}\right) +\nabla \times \boldsymbol{M}\left( r\right) \right] ,
\end{equation}%
where%
\begin{equation*}
\boldsymbol{v}_{n}\left( \boldsymbol{k}\right) =\frac{\partial \varepsilon
_{n}\left( \boldsymbol{k}\right) }{\hbar \partial \boldsymbol{k}}+\dot{F}%
_{\perp }\boldsymbol{\Omega }_{n,F_{\perp }\boldsymbol{k}}-\frac{\boldsymbol{%
F}_{\parallel }}{\hbar }\times \Omega _{n,\boldsymbol{k}}\hat{z},
\end{equation*}%
is the velocity of an exciton in a bilayer system derived from
semi-classical theory \cite{Xiao,Xiao1,Xiao2}. The velocity consists of
three parts: the band velocity, the anomalous velocity induced by hybrid
Berry curvature $\boldsymbol{\Omega }_{n,F_{\perp }\boldsymbol{k}}$ in the $%
\left( F_{\perp },\boldsymbol{k}\right) $ space, and by the $\boldsymbol{k}$%
-space Berry curvature $\Omega _{n,\boldsymbol{k}}$. Here $F_{\perp
}=F_{\perp }^{0}f\left( t\right) $ is an out-of-plane mechanical force that
may arises from the detuning of exciton resonance caused by layer dependent
screening, with $f\left( t\right) $ being a function dependent on time $t$. $%
\boldsymbol{F}_{\parallel }=\hbar \frac{d\boldsymbol{k}}{dt}$ represents the
in-plane general mechanical force, where%
\begin{equation*}
\boldsymbol{M}\left( r\right) =\frac{k_{B}T}{\hbar }\sum_{n}\int \frac{d^{2}%
\boldsymbol{k}}{\left( 2\pi \right) ^{2}}\Omega _{n,\boldsymbol{k}}\log
\left( 1-e^{-\left( \varepsilon _{n}-\mu \right) /k_{B}T}\right) ,
\end{equation*}%
is analogous to the equilibrium magnetization density in electronic systems 
\cite{Wang,Di}.

The semi-classical theory and symmetry analysis finally gives \cite{Supp} 
\begin{eqnarray}
\boldsymbol{j} &\equiv &\boldsymbol{j}_{H}=\boldsymbol{j}^{\omega }\sin
\left( \omega t\right) ,  \label{CND} \\
\boldsymbol{j}^{\omega } &=&\boldsymbol{F}_{\parallel }\times \sigma _{H}%
\hat{z}+k_{B}\nabla T\times \alpha _{H}\hat{z},  \notag
\end{eqnarray}%
under an assumption that $F_{\perp }=F_{\perp }^{0}\cos \left( \omega
t\right) $. The schematic illustration is shown in Fig. \ref{fig4}(a). $%
\sigma _{H}=\omega F_{\perp }^{0}\chi ^{\mathrm{int}}$ is the crossed
nonlinear dynamical Hall conductivity and $\alpha _{H}=\omega F_{\perp
}^{0}\chi _{\mathrm{Ner}}^{\mathrm{int}}$ represents the crossed nonlinear
dynamical Nernst conductivity, where%
\begin{eqnarray*}
\chi ^{\mathrm{int}} &=&\frac{1}{\hbar }\sum_{n}\int \frac{d^{2}\boldsymbol{k%
}}{\left( 2\pi \right) ^{2}}f_{0}^{n}\left( \nabla _{\boldsymbol{k}}\times 
\boldsymbol{\mathcal{G}}^{n}\left( \boldsymbol{k}\right) \right) _{z}, \\
\chi _{\mathrm{Ner}}^{\mathrm{int}} &=&\frac{1}{\hbar }\sum_{n}\int \frac{%
d^{2}\boldsymbol{k}}{\left( 2\pi \right) ^{2}}f_{0}^{n}\left( \nabla _{%
\boldsymbol{k}}\times \boldsymbol{\mathcal{G}}^{n}\left( \boldsymbol{k}%
\right) \right) _{z}\cdot  \\
&&\left[ \frac{\varepsilon _{n}-\mu }{k_{B}T}f_{n}^{0}-\log \left(
f_{n}^{0}\exp \left( \frac{\varepsilon _{n}-\mu }{k_{B}T}\right) \right) %
\right] ,
\end{eqnarray*}%
is intrinsic to the band structure. Here, the unusual quantum geometric
properties induced by the chiral symmetric layer hybridized wavefunctions is
measured by the $\boldsymbol{k}$-space curl of the interlayer Berry
connection polarizability (BCP)%
\begin{equation*}
\boldsymbol{\mathcal{G}}^{n}\left( \boldsymbol{k}\right) =2\hbar ^{2}\mathrm{%
Re}\sum_{m\neq n}\frac{\left\langle u_{n}\right\vert p_{z}\left\vert
u_{m}\right\rangle }{\left( \varepsilon _{n}-\varepsilon _{m}\right) ^{3}}%
\boldsymbol{v}_{mn},
\end{equation*}%
over the occupied states \cite{Xiao,JC}. Here, $p_{z}=\Delta z\hat{\sigma}%
_{z}$ is the interlayer dipole moment. Although CNDHE remains forbidden in
TMDs without strain since the rotational symmetry of the related exciton
Hamiltonian, it can emerge after applying strain, as demonstrated in Fig. %
\ref{fig4}(b), which shows a similar dependence to that observed in the
TREHCF calculation. This behavior is dominated by the quantum geometric
properties of $\left( \nabla _{\boldsymbol{k}}\times \boldsymbol{\mathcal{G}}%
^{n}\left( \boldsymbol{k}\right) \right) _{z}$, as shown in Fig. \ref{fig3}%
(c) \cite{Note1}. In BP, however, a significant value of CNDHE appears at
specific twisted angle $\theta $ (Fig. \ref{fig4}(c)), corresponding to the
distribution of BCP with respect to $\theta $, as shown in Fig. \ref{fig3}%
(d). The periodicity dependent on $\theta $ is $\pi $ since only $C_{2}$
symmetry are left in $H_{\mathrm{BP}}$. Compared to the TREHCF, the CNDHE is
more dependent on the material's anisotropy and the specific twisted angle $%
\theta $, as evidenced by the numerical calculations in Fig. \ref{fig3}(d).
This finding may suggest new opportunities for applications of BP in
exciton-based optoelectronics.

\begin{acknowledgments}
C. L. would like to thank D. W. Zhai, B. Fu, C. Xiao, and B. B for useful discussions.
This work is supported by the National Key R$\&$D Program of China
(2020YFA0309600), Research Grant Council of Hong Kong SAR (AoE/P-701/20, HKU SRFS2122-7S05), and New Cornerstone Science Foundation.
\end{acknowledgments}

\end{document}